\documentstyle[11pt,newpasp,twoside,epsf]{article}
\def\edcomment#1{\iffalse\marginpar{\raggedright\sl#1\/}\else\relax\fi}
\reversemarginpar

\begin{document}
\title{Gamma-Ray Bursts environment and progenitors}
 \author{Davide Lazzati}
\affil{Institute of Astronomy, University of Cambridge, Madingley Road, 
CB3~0HA Cambridge (UK)}

\begin{abstract}
The properties of the ambient medium in which GRBs go off are an
important piece of the puzzle, not only as an issue in itself, but
because of their link with the nature of the progenitor. In this
review, I describe and critically comment the various experiments
proposed to study the burst environment. I discuss emission and
absorption features techniques and compare their results to afterglow
modelling. A consistent picture cannot be drawn, and I suggest that
the evaporation of the soft X-ray absorbing column is the more
promising of the tools, even though the fast reaction of Swift is
required in order to obtain data of sufficiently high quality.
\end{abstract}

\section{Introduction}

To understand the properties of the environment of GRBs is of great
importance for at least three reasons. First, modelling the afterglow
emission in a known environment would allow us to better understand
the properties of collisionless shocks; second, it is of great
interest to understand how the ISM reacts to the huge ionizing flux of
the burst and its afterglow, that ionizes the gas and sublimates dust
grains.  Finally, and most importantly, different progenitors are
expected to be surrounded by sizably different environments, and
therefore knowing the environment of GRBs means identifying their
progenitors.

I will discuss in this paper three different classes of progenitors
which, to some extent, encompass most of the dozens different
varieties of progenitors proposed in the literature so far. These
three will be called i) binary mergers; ii) hypernov\ae~and iii)
supranov\ae.

The first class contains all the progenitor models in which the burst
is caused by the collapse of a binary system, in which one or both the
components are degenerate. The familiar case is that of binary neutron
stars (Eichler et al. 1989).  This class predicts a uniform and low
density environment, since binary systems travel far from their
birth-place before merging. The burst should explode outside of the
star formation cocoon where the binary system was born. SN-SN systems
would collapse in the host galaxy ($1\le n\le100$~cm$^{-3}$, while
SN-BH would explode in the IGM.  ($n\le1$~cm$^{-3}$; Perna
\& Belczynski 2002)

The second class is that of hypernov\ae~(MacFadyen, Woosley, \& Heger
2001), now most popular. Here the burst is supposed to explode at the
end of the life of a massive, possibly rotating, star which has gone
through a phase of heavy wind ejection, in order to get rid of its
hydrogen -- and possibly helium -- envelope. Such a progenitor, at the
moment of the burst explosion, should be surrounded by a complex
environment. At small radii (up to $\sim0.01$~pc) the pre-explosion
stellar wind dominates, with a density scaling with the radius as
$n\propto{}R^{-2}$. At larger distances, a rather sharp transition
connects the wind with the more uniform density provided by the
molecular cloud. A large and uniform density
($10^2\le{}n\le10^4$~cm$^{-3}$) is then expected at large radii.

The properties of the environment of supranov\ae~are easily the most
complex of the lot. In supranov\ae, the GRB explosion takes place
inside a relatively young supernova remnant (SNR), generated by the
explosion of a supernova (SN) that took place several months to
several years prior to the GRB (Vietri \& Stella 1998). The close
environment of the burst progenitor is therefore the inside of a SNR,
possibly modified by the emission of a highly magnetized pulsar
(K\"onigl \& Granot 2002), and likely elongated along the rotational
axis of the progenitor star. At the edge of the SNR, a condensation of
matter is expected in the form of a shell, possibly clumped similarly
to the Crab SNR. The distance of this shell from the burst explosion
site depends on the time delay between the SN and GRB explosions and
can influence the fireball dynamics as well as the reprocessing of the
burst radiation (see below).

\section{Emission Features}

Narrow emission features in the X-ray afterglow of GRBs have been
claimed in at least five events to date (see, for a review, Lazzati
2002 and references therein). Similarly to the case of AGNs, the
presence and evolution of emission features can be used to map the
environment of the burst itself with the reverberation mapping
technique. Its application, even if in principle extremely powerful,
requires however a data quality which is far beyond what we can hope
to have now and in the near future in GRBs. We have in fact to deal
with moderately significant detections (still under heavy debate and
criticism), and it is not conceivable to measure the evolution of the
intensity of these features.

The presence of the lines is not, however, meaningless. Lines with the
detected properties can not in fact be produced in a low density
environment\footnote{Note that I am discussing here the fireball
model. The cannonball model can explain the features in a different
way. See Dado, Dar, \& De Rujula (2002) for details.}. Consider an
iron $K_\alpha$ line with $L_{\rm{line}}\sim10^{44}$~erg~s$^{-1}$
observed for a time scale $t_{\rm{line}}\sim1$~day. Since the line
emission cannot be beamed (Ghisellini et al. 2002) the total number of
line photons produced is
$N_{\rm{line}}=L_{\rm{line}}\,t_{\rm{line}}/(h\nu_{\rm{line}})\ga10^{57}$. In
a low density environment, recombination is not effective, and this
implies a total number of iron atoms $N_{\rm{Fe}}\ga10^{56}$; since an
isolated iron ion can produce up to $\sim10$ $K_\alpha$ line photons
due to Auger auto-ionization. Considering light-travel effects, one
can set a limit to the size of the emitting region by requiring that
the total energy in line photons is not larger than the total energy
in the burst itself:
$R_{\rm{line}}\la{}c\,t_{\rm{line}}\la{}c\,E_{\rm{line}}/L_{\rm{line}}\la10^{17}$~cm,
where a total energy in line photons $E_{\rm{line}}\la10^{51}$~erg has
been assumed. This implies $n_H\ga10^{7}$~cm$^{-3}$.

The importance of this limit is in the fact that the fireball cannot
be propagating in such a large density, since a non relativistic
transition would take place at $\sim1$~day. Such an early transition
has never been observed (with the possible exception of GRB990705,
Lazzati et al. 2001). Explaining this apparent contradiction is the
riddle to be solved in order to explain the origin of the lines. Two
possible solutions have been found, to date.

\begin{figure}
\plotone{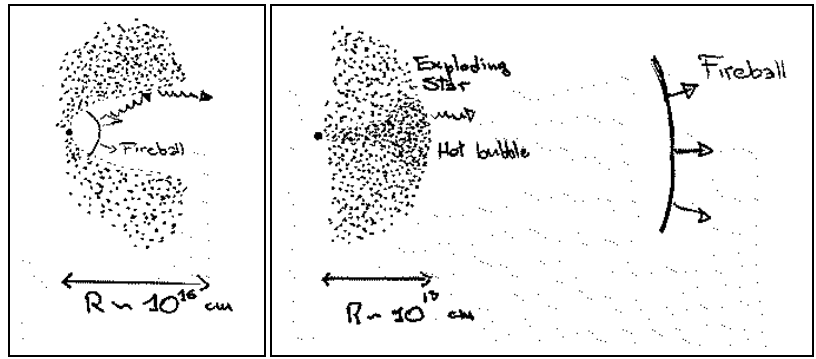}
\caption{Cartoon showing the basic features of the two models 
described in the text. The left panel shows a Geometry Dominated
model while the right panel shows an Engine Dominated case.}
\end{figure}

\subsection{Geometry dominated mechanisms}

The two solutions to the high density problem outlined above can be
identified through the way in which the duration of the line is
explained. In the first case the line is thought to be produced in an
extended region surrounding the burst progenitor, whose size is large
enough to make the line emission observable for $\sim1$~day due to
light-travel effects. This solution is called ``Geometry Dominated''
(GD; see left panel of Fig.~1). In order to avoid the density problem,
the high density material has to be asymmetrically distributed.
Examples of such an asymmetric distribution are an elliptical, pierced
of funnel-like SNR (Lazzati, Campana, \& Ghisellini 1999; Vietri et
al. 2001) or an extended disk or torus surrounding the progenitor
(B\"ottcher 2000). The continuum radiation that provides the ionizing
photons is that of the burst and afterglow itself (Lazzati,
Ramirez-Ruiz, \& Rees 2002a). This line mechanisms is associated to
supranov\ae, since the SN explosion naturally provides the high
density material at the correct distance from the burster.

\subsection{Engine dominated mechanisms}

An alternative solution to the density problem can be envisaged if the
high density material is confined in a very small and dense region,
which is left behind by the fireball at a very early stage (Rees \&
Meszaros 2000; see the right panel of Fig.~1). In this case, however,
the emitting material cannot be illuminated by the GRB and afterglow,
and a tail of emission from the inner engine has to be postulated. For
this reason, the line emission lasts as long as the engine is active
and this class of models is called ``Engine Dominated'' (ED). Line
production in ED scenario is associated to the hypernova progenitor
class, since the dying star naturally provides the compact high
density material.

\subsection{GD or ED?}

It is therefore clear that an indication of the progenitor nature
would come by understanding which of the two scenarii for line
emission is correct. Without invoking extremely accurate datasets,
some key observations hopefully performed in the near future should be
able to clarify the matter. In the GD-supranova scenario, in fact, the
observed line properties strongly depend on the time delay between the
SN and GRB explosions. In particular, the detection of Co lines is
possible only in this case and should be at some point observed, for a
GRB-SN delay $20\la{}t_\Delta\la100$~d. In addition, the duration of
the line emission is in this model related to $t_\Delta$, and
therefore, a correlation between the line variability and the line
identifications should be possible. Presently, the sparse evidences
point to a GD scenario (Lazzati et al. 2002a), but more accurate
measurements are required in order to draw any strong conclusion.

\section{Absorption Features}

In alternative to the detection of emission features, the detection
and characterization of absorption features has been proposed as a
tool to investigate the burst environment. Absorption by itself,
however, is sensitive to the column density ($N_H$) along the line of
sight, and not to the local density, which is the goal of our
discussion. This limitation can be overcome if the absorption process
is destructive. If, in fact, the absorption of a photon causes the
destruction of the absorber, the opacity of the intervening medium
decreases in time, at a speed which is faster the closer is the
absorber to the GRB. Detecting a variable absorption can therefore
give us precious informations on the radial density and structure of
the medium surrounding the GRB.

Different absorption effects have been suggested as a tool: resonant
optical lines (Perna \& Loeb 1998), X-ray photoionization edges
(Lazzati, Perna, \& Ghisellini 2001), X-ray continuum absorption
(Lazzati \& Perna 2002) and dust extinction (Waxman \& Draine 2000;
Perna, Lazzati, \& Fiore 2003). Let us discuss them starting from the
lowest frequencies.

\subsection{Dust destruction}

Suppose the burst progenitor, as natural in the hyper- and supranova
scenario, explodes within a molecular cloud or, generally speaking, a
dense region of gas and dust grains. The dust grains will absorb near
UV light from the burst and afterglow, as well as X-ray photons, and
will be heated to large $\ga3000$~K temperatures as a result. The
grain will also cool by black-body radiation and, if this is not
enough to keep the temperature low, by sublimation of its external
layers. As a result, the grain will be completely dissolved after a
certain lapse of time, which is smaller than the burst duration out to
distances of the order of several parsecs. An important characteristic
of this process is that the sublimation rate is sensitive to the
optical flux (and not to the fluence) and therefore most of the dust
evaporation takes place during the most luminous phase of the GRB
in the optical: the optical flash. For this reason, detecting a
reddening evolution is extremely hard. It requires multiwavelength
observations at times smaller than the peak time of the optical flash
which, for the only case studied so far (albeit with unfiltered
images), is as short as $T_{\rm{OF}}\sim40$~s. This may explain why,
even if reddening and dust destruction have been claimed for many GRB
afterglows, this phenomenon has never been caught ``on the act''.

An alternative dust destruction process sensitive to the X-ray fluence
has been recently proposed. Is is based on the Coulomb Explosion (CE)
effect (Fruchter, Krolik, \& Rhoads 2001). This effect is in
competition with the Ion Field Emission (IFE; see e.g. Perna \&
Lazzati 2002), which is also based on X-ray photoionization of the
grain and can quench the CE completely. It is not clear whether CE or
IFE dominate, but their observable consequences are different. Should
any reddening evolution be detected during the afterglow phase, it
would certainly point to a fluence sensitive mechanism and therefore
confirm the role of CE in the grain sublimation. This could give
important insight in our understanding of the structure of cosmic dust
at high redshift.

\begin{figure}
\plottwo{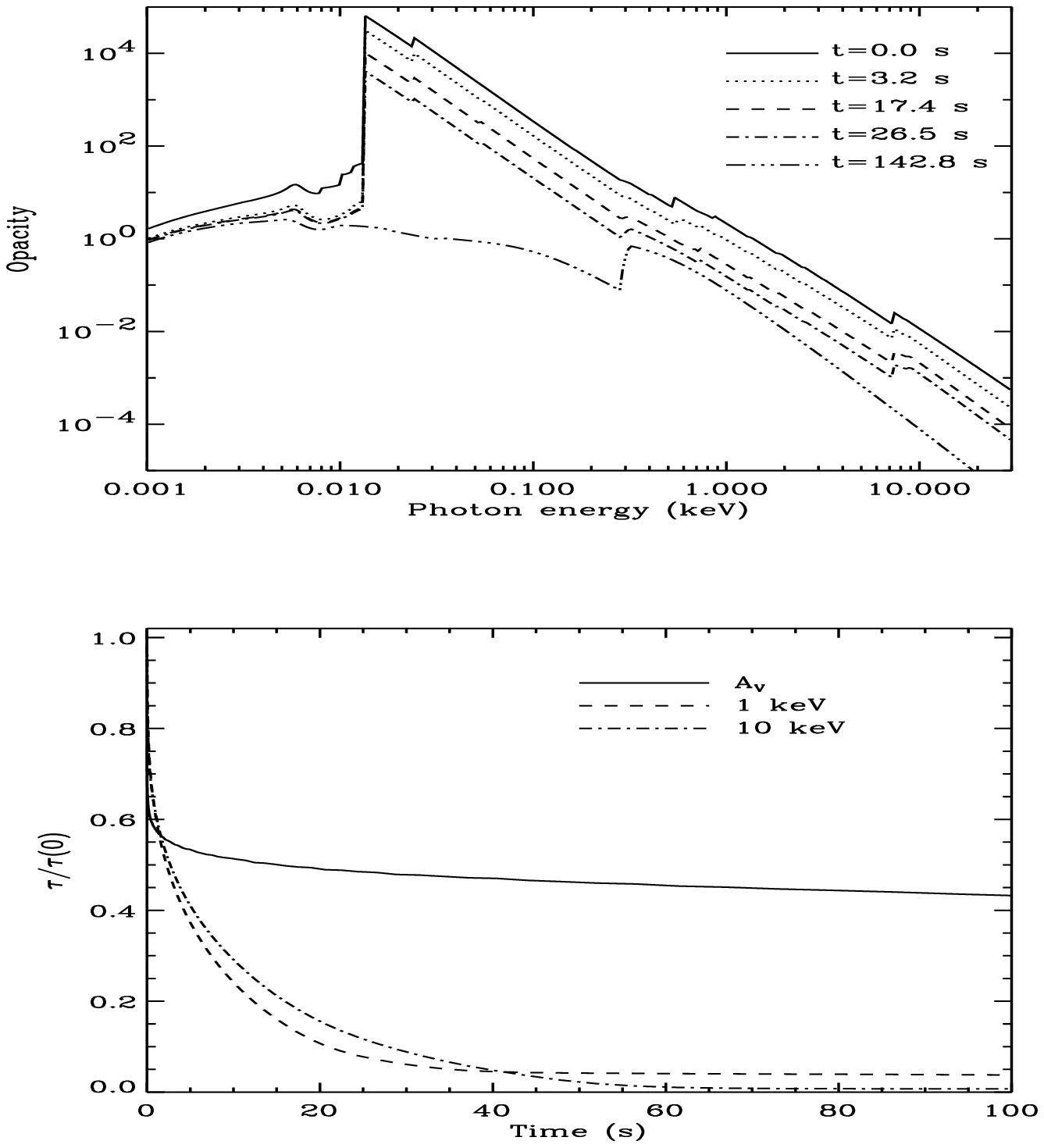}{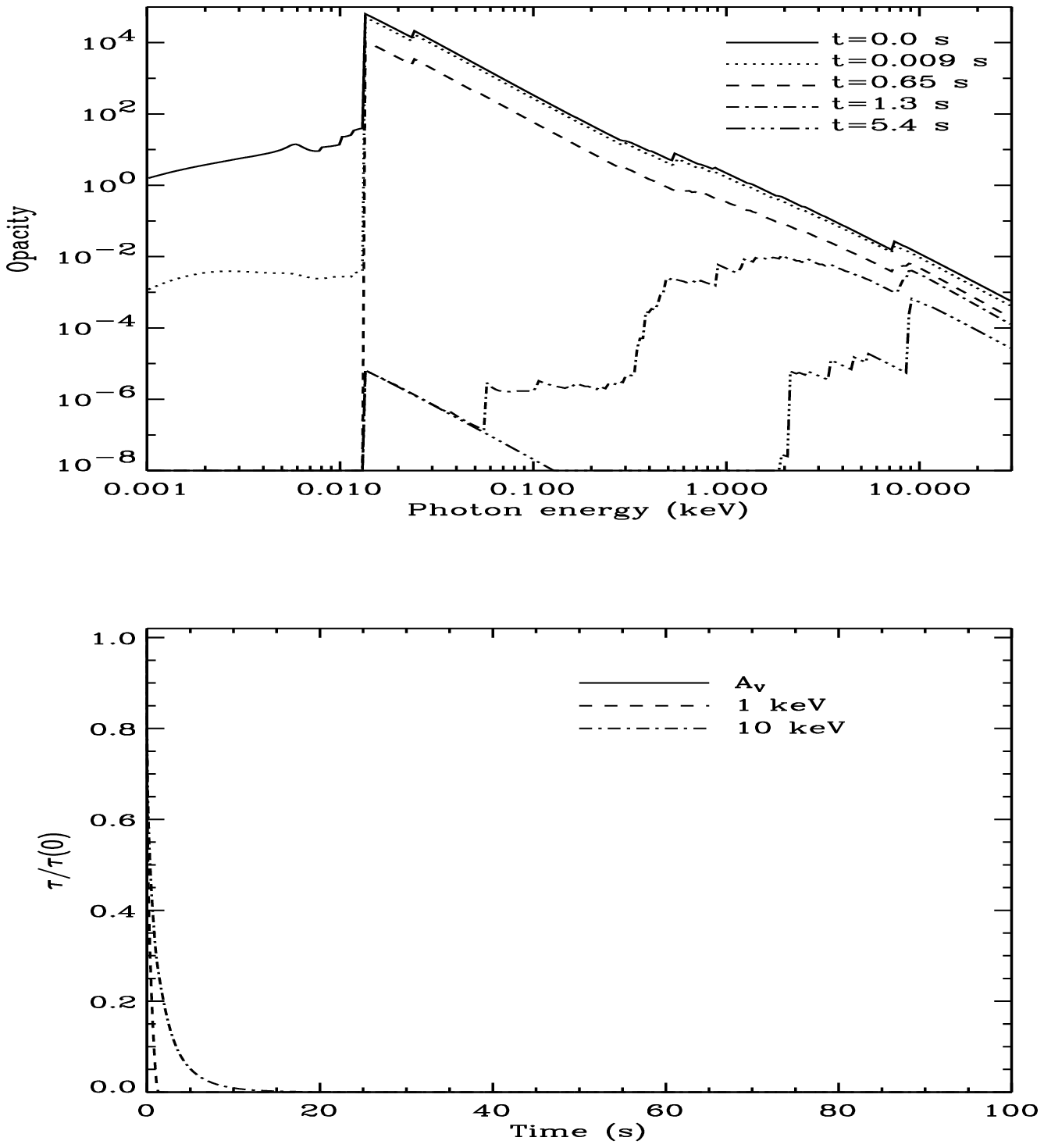}
\caption{Evolution of the opacity at various frequencies for two 
simulations. In the left panels a GRB without optical flash is
exemplified by a spectrum $L(\nu)\propto\nu^0$, while in the right
panels a GRB with optical flash has $L(\nu)\propto\nu^{-1/2}$.}
\end{figure}

\begin{figure}
\plotone{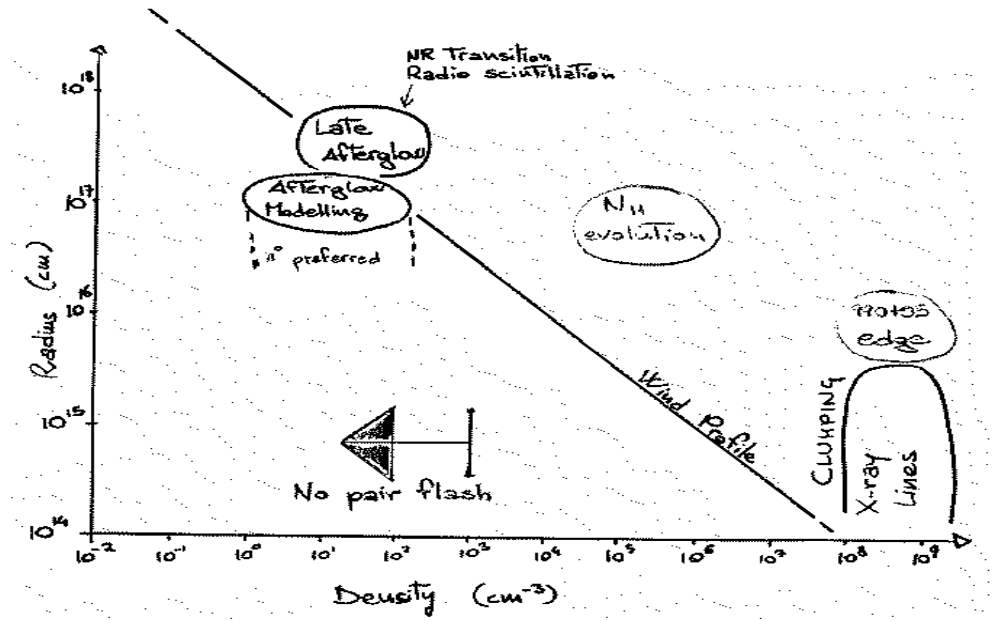}
\caption{Summary of the density measurements performed with various 
techniques plotted versus the distance from the GRB site at which they
are applicable.}
\end{figure}

\subsection{Photoionization}

The other process that can influence the opacity of the intervening
medium is photoionization. The evolution of the opacity due to
photoionization can be observed in the optical as a reduction of the
equivalent width (EW) of resonant lines and in the X-rays as reduction
of the continuum absorption or of the opacity of a single
photoionization edge. Since photoionization, for any reasonable
density of the intervening medium, is a destructive process
(recombination times are much longer than several days or weeks) and
is sensitive to the photon fluence, its effects can be observed easily
during the afterglow phase.

It is extremely important to stress that in order to understand any
possible observation (or non observation) of opacity evolution, an
accurate knowledge of the spectral fluence of the whole GRB+afterglow
phenomenon up to the moment of the observed evolution is required. It
is not possible to constrain the properties of the absorber with the
mere knowledge of the afterglow flux at the moment in which the
opacity evolution was detected. This is the main reason why I consider
soft X-ray opacity evolution (or $N_H$ evolution) my best bet as a
probe of the GRB environment. In fact, it is much easier to know the
fluence of a GRB in the X-rays than it is in the optical and therefore
to constrain the amount of evaporation already occurred in the X-rays
than in the optical band.

In order to translate possible evidences of opacity evolution into
properties of the ambient medium, time dependent simulations are
required. As it has been shown by Lazzati \& Perna (2003) such
simulations must include all the processes discussed above, since
their effect intertwines and the result depends on the superposition of
them. In Fig.~2 I show the results of two simulations from the code of
Perna \& Lazzati (2002), for a spherical uniform absorbing cloud of
radius $R=10^{18}$~cm. The average luminosity of the burst in the
[1~eV--100~keV] range was $L=10^{50}$~erg~s$^{-1}$ and the cloud
initial column density was $N_H=10^{22}$~cm$^{-2}$, with a standard
dust-to-gas ratio. What can be drawn by comparing the left and right
panels of the two figures is that i) the rate of dust sublimation
depends strongly on the spectral shape of the prompt emission and ii)
the decrease rate of X-ray opacity is less sensitive to it.

\subsection{Results}

Variable absorption techniques have been applied to a handful of GRB
and afterglow measurements to obtain constraints on the cloud
properties. In the optical Mirabal et al. (2002), using the constancy
of the EW of optical lines in two spectra of GRB~010222, derived some
constraints on its environment. In the X-rays, Lazzati \& Perna (2002)
constrain the environment of two GRBs by modelling the possible (a
$3\sigma$ effect) evolution of soft X-ray absorption during their
prompt emission, while Lazzati et al. (2001) derived the size, density
and clumpiness of the surroundings of GRB~990705 modelling the
transient feature in its prompt X-ray spectrum (Amati et
al. 2000). All these X-ray measurements constrain the ambient medium
to be compact (a fraction of a parsec) and fairly dense (up to
densities $n\sim10^{10}$~cm$^{-3}$ in the clumps).

\section{Afterglow modelling}

The variable absorption techniques discussed above yield a model
independent estimate of the ambient density and profile, but are
hampered by the lack of accurate measurements of variable
absorptions. An alternative way of measuring densities and density
profiles for afterglows is provided by the modelling of the afterglow
emission itself. In fact, the intensity and position of the spectral
breaks of an afterglow, coupled with their temporal evolution, provide
us with enough information to constrain most of the afterglow
parameters, included the density and density profiles of the ambient
medium. Such an estimate can be made with a sufficiently large number
of data points of great quality. However, it is highly model
dependent, and any numerical value is precise as long as the fireball
model is considered accurate. Limitation of the model are its
extremely simplified treatment of the magnetic field generation and
propagation (Rossi \& Rees 2003) and the constancy of the
equipartition parameters.

\section{Summary and Conclusions}

Figure~3 shows a collection of measurements of densities of the
environment as a function of the distance from the GRB site at which
the measure is applicable. To date is not possible to draw an unique
conclusion from the collection of all measurements. In fact,
measurement derived from the modelling of the afterglow lightcurves
suggest a low density uniform medium (Panaitescu \& Kumar 2002)
possibly with moderate clumping (Lazzati et al. 2002b). They are
applicable to large distances ($\sim1$~pc), but the indication of
uniformity allow us to extend their validity to smaller
scales. Consistent measures are obtained by late measurement of the
fireball evolution, such as relativistic non-relativistic transitions
and radio calorimetry (Frail, Waxman, \& Kulkarni 2000).

On the other hand, all the model independent measures of density,
based on emission lines and/or transient absorption features, yield
environments constrained to be dense and highly clumped at small
scales (tenth of pc).

Both methods have their pros and cons. Methods based on afterglow
modelling are intrinsically model dependent and therefore require an
independent assessment of their validity. On the other hand, methods
based on variable absorption are now applied to low quality datasets,
and therefore the obtained results are affected by large
uncertainties, both statistical and systematic.

With the advent of future missions and the increased quality of the
available data, we can foresee that the latter methods will become
more and more reliable, and hopefully in the near future we will be
able to constrain the GRB environment with absorption measures, plug
these results in the afterglow modelling and use the afterglow
measurements as a formidable laboratory for the shock physics and
hydrodynamics.

\acknowledgements I am grateful to Gabriele Ghisellini, Rosalba Perna, 
Luigi Stella and Mario Vietri for their contributions to some of the
work presented in this paper.


\begin{references}
\reference Amati, L.~et al.\ 2000, Science, Volume 290, Issue 5493, 
	pp.~953-955 (2000)., 290, 953
\reference Dado, S., Dar, A.~\& De Rujula, A. 2002, ApJ in 
	press (astro-ph/0207015)
\reference Eichler, D., Livio, M., Piran, T., \& Schramm, D.~N.\ 1989, 
	Nature, 340, 126
\reference Frail, D.~A., Waxman, E., \& Kulkarni, S.~R.\ 2000, \apj, 537, 191
\reference Fruchter, A., Krolik, J.~H., \& Rhoads, J.~E.\ 2001, \apj, 563, 597 
\reference Ghisellini, G., Lazzati, D., Rossi, E., \& Rees, 
	M.~J.\ 2002, \aap, 389, L33
\reference K{\" o}nigl, A.~\& Granot, J.\ 2002, \apj, 574, 134
\reference Lazzati, D., Campana, S., \& Ghisellini, G.\ 1999, \mnras, 304, L31
\reference Lazzati, D., Ghisellini, G., Amati, L., Frontera, F., 
	Vietri, M., \& Stella, L.\ 2001, \apj, 556, 471
\reference Lazzati, D., Perna, R., \& Ghisellini, G.\ 2001, \mnras, 325, L19 
\reference Lazzati, D., 2002, to be published in Beaming and Jets in 
Gamma Ray Bursts, ed. R. Ouyed, J. Hjorth \& \AA. Nordlund (astro-ph/0211174)
\reference Lazzati, D.~\& Perna, R.\ 2002, \mnras, 330, 383 
\reference Lazzati, D., Ramirez-Ruiz, E., \& Rees, M.~J.\ 2002a, \apjl, 572, L57
\reference Lazzati, D., Rossi, E., Covino, S., Ghisellini, G., \& 
	Malesani, D.\ 2002b, \aap, 396, L5
\reference Lazzati, D. \& Perna, R., 2003, MNRAS in press (astro-ph/0212105)
\reference MacFadyen, A.~I., Woosley, S.~E., \& Heger, A.\ 2001, \apj, 550, 410
\reference Mirabal, N.~et al.\ 2002, \apj, 578, 818 
\reference Panaitescu, A.~\& Kumar, P.\ 2002, \apj, 571, 779
\reference Perna, R.~\& Loeb, A.\ 1998, \apj, 501, 467
\reference Perna, R.~\& Belczynski, K.\ 2002, \apj, 570, 252 
\reference Perna, R.~\& Lazzati, D.\ 2002, \apj, 580, 261 
\reference Perna, R., Lazzati, D., \& Fiore, F., 2003, ApJ in press 
	(astro-ph/0211235)
\reference Rees, M.~J.~\& M{\' e}sz{\' a}ros, P.\ 2000, \apjl, 545, L73 
\reference Vietri, M.~\& Stella, L.\ 1998, \apjl, 507, L45 
\reference Vietri, M., Ghisellini, G., Lazzati, D., Fiore, F., \& 
	Stella, L.\ 2001, \apjl, 550, L43
\reference Waxman, E.~\& Draine, B.~T.\ 2000, \apj, 537, 796 

\end{references}
\end{document}